\documentclass[prl,twocolumn,aps,superscriptaddress,showpacs,floatfix]{revtex4}
\usepackage{amsmath}
\usepackage{graphicx}

\begin{document}
\title{Energy dissipation in small scale shape-change dynamics}

\author{L. Gammaitoni\cite{lgemail} }
\address{NiPS Laboratory, Dipartimento di Fisica, Universit\'a di Perugia, and Instituto 
Nazionale di Fisica Nucleare, Sezione di Perugia, I-06100 Perugia, Italy}

\date{\today}

\begin{abstract}
Shape is an important feature of physical systems although very seldom it is addressed in the framework of a quantitative description approach. In this paper we propose to interpret the shape of things as a physical manifestation of the content of information associated with each thing and show that a change of shape in a physical system is necessarily connected with a change of its entropy and thus involves energy. We estimate the amount of energy dissipated during a shape change and propose experimental tests to be performed in nanoscale systems, to verify this prediction by measuring the expected dissipation in few simple cases. Relevant implications in the design of future zero-power logic switches are discussed.
\end{abstract}
\pacs{65.40.gd, 89.70.Cf, 05.70.-a, 05.10.Gg, 05.40.-a}

\maketitle

In a world where things are made by tiny particles, the shape of things at finite temperature changes spontaneously with time, according to a diffusion process that is a manifestation of the second principle of thermodynamics. A typical example would be the shape of airplane contrails. In this case the initial positions of the particles of condensed water vapor generated by the exhaust of aircraft engines fit in a straight narrow line. As time passes however, the line gets smeared and eventually disappears. Other examples are the change in shape of ink drops in a water bowl or the dramatic change of shape of concrete buildings during an earthquake. 
All these cases are examples of physical phenomena whose main macroscopic aspect is the change of shape.

\begin{figure}[b]
\includegraphics*[width=9cm]{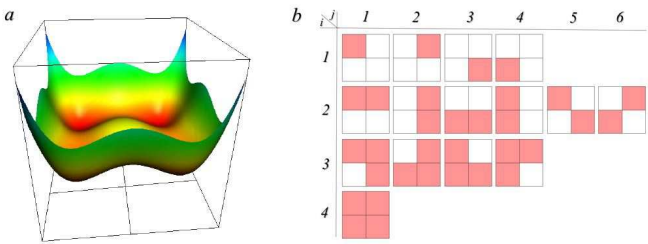}
\caption{Physical system and its schematization. Left: confining potential energy landscape with four minima sites each representing a dynamically stable equilibrium point for a material particle. Right: Four particles can occupy up to four sites. The site occupation is signaled by a colored square in the four-squares scheme sketched under the potential. In this table, the rows indicate shape classes (index $i$) while columns indicate different shapes within the same class (index $j$).
}
\label{F1} 
\end{figure}

In this letter we address the following question: what is the minimum energy required to change shape to a physical object? We will show that the shape of an object is a manifestation of the content of information associated with that object and establish a quantitative connection between the change of its shape and the minimum energy dissipation required by such a change.

To reach this goal we start with introducing the physical observable {\em shape entropy} by means of a toy model that mimics the relevant features of the shape change dynamics of a physical object. Subsequently we proceed by establishing the connection between the {\em shape entropy} change associated with a shape change and the minimum energy requirement. We discuss the implication of this result in the study of low dissipation physical systems with specific reference to the design of zero-power logic switches, the building block of future ICT devices. Finally we propose few simple experimental tests to be performed in nano scale systems, to verify our predictions by measuring the resulting energy dissipation in few selected cases.

In order to introduce the physical observable {\em shape entropy} $S$ we consider a toy model representing an highly idealized physical object. Let's suppose our physical object is made by a small number of elementary components or parts that we will generically call particles. The shape of this object is the result of the way the constituting particles are arranged in the physical environment occupied by our object. To fix our ideas let's consider a simple object made by four indistinguishable particles that can be arranged in a 2x2 sites (see Figure 1a) represented by 4 potential energy minima. The occupancy of a site is signaled by a corresponding pink square underneath Figure 1a. Empty sites are represented by white squares. Now we make the oversimplified assumption that our four-particle object can assume different shapes according to the different dispositions of the four particles in the four sites, based on the simple rule: one or more particles can occupy the same site at the same time with equal probability. Based on these premises we can easily count the different shapes generated by the different dispositions of the four particles (Figure 1b). We can group the different shapes in {\em classes}: all the shapes that have the same number of occupied sites belong to the same class. A given shape can be realized by a number of different {\em configurations}, i.e. different dispositions of the four particles. 

In general if we have $q$ indistinguishable particles that can be distributed in $r$ distinguishable sites, a single shape $s_{ij}$ is characterized by two indexes: the class index $i=1,2,..r$  and, within a single class, the shape index $j=1,2,..C(r,i)$ where $C(r,i)$ is the binomial coefficient. The total number of different shapes is given by 

\begin{equation}
\label{NShapes}
N_S=\sum_{i=1}^{r}{C(r,i)}
\end{equation}

The number of configurations for each given shape $s_{ij}$, $N_{ij}$,  depends only on the shape class, i.e. $N_{ij}=N_i$ and this is given by:

\begin{equation}
\label{Ni}
N_{i}=C(q-1,i-1)={\frac{(q-1)!}{(i-1)! (q-i)!}}
\end{equation}

The total number of possible configurations is given by $N=C(q+r-1,r-1)$.

In our example with $q=4$ and $r=4$ we have $N_S=15$ and $N=35$ while $N_1=1$, $N_2=3$, $N_3=3$ and $N_4=1$.

In order to gain some insight on how the shapes change one into another we apply a random shaking to our system of particles. The shape change is produced by one or more particles that switch site by crossing the potential barrier between the sites, under the action of the random force. 
A proper treatment of this kind of problems requires a stochastic dynamic approach\cite{1}. However, without carrying out a detailed calculation, we can summarize here few relevant aspects: i) The asymptotic stationary probability density function for the shape distribution $P_{ij}$ (where $P_{ij}$ is the probability of observing a shape $s_{ij}$) exists and is reached after a transient phase whose details depend on the specific form of the energy landscape. ii) $P_{ij} =P_{i}$ i.e. the probability inside a class is the same for all the shapes that belong to that class. iii) $P_{i} = N_i/N$, i.e. the shapes that have the largest number of possible configurations are the shapes with the largest probability of realizations under the action of the random shaking. 
We define {\em shape entropy} the quantity
\begin{equation}
\label{Shape-e}
S_{i}=K \ln N_i
\end{equation}

where $K$ is an arbitrary constant.
This quantity coincides with the microscopic form given by Boltzmann and Gibbs of the thermodynamic entropy initially introduced by Clausius, if we interpret the number of configurations $N_i$ for a given shape as the number of accessible microstates for a given state of the thermodynamical system.
Specifically, Gibbs entropy is given by

\begin{equation}
\label{Gibbs-e}
S_G = -K \sum_l{p_l \ln{p_l}}
\end{equation}

where $p_l$ is the probability of the microstate of index $l$ and the sum is taken over all the microstates. This expression reduces easily to (\ref{Shape-e}) (that is formally identical to the Boltzmann entropy if $K = k_B$)  when all the $p_l$ are equal as in our example of the four particles. 

Thus, if our particle system can be considered at thermal equilibrium at a certain temperature $T$, then the random shaking will be provided by its thermal energy and its dynamical evolution will be subjected to the second law of thermodynamics that requires that the system spontaneously evolves toward the maximization of its entropy. 

Up to this point we have shown that the shape of a physical object can be associated with a physical observable called "shape entropy" and that this object in thermal equilibrium with the environment will spontaneously change its shape according to the maximization of the shape entropy.

Now we take one step further and show that the shape entropy is indeed a measure of the quantity of information associated with the shape of an object. The connection between quantity of information and shape entropy is readily established on the base of the formal analogy between the microscopic entropy by Gibbs and the expression introduced by Shannon\cite{8} to quantify the information content of a given message chosen from a set of all possible messages. This is called the Shannon entropy and it is formally identical to (\ref{Gibbs-e}) where $p_l$ represents the probability of receiving the message with index $l$ and $K$ is a constant that fixes the units. Based on this analogy we can compute the quantity of information a la Shannon for the different shapes in our model. In order to associate an information content to a shape we select the following coding system. We use 2 bits per site identifying the occupation of a site as follows: a particle on the upper left is characterized by $00$, upper right $01$, lower right $10$, lower left $11$. One configuration is represented by the occupation of the fours sites and thus requires 8 $bits$ (whose order is immaterial due to undistinguishable character of the particles).
The question we want to answer is the following: how much information is contained in a shape represented by a certain group of binary digits? The answer is promptly obtained by computing the Gibbs entropy in the Shannon interpretation. As we have seen a given shape can be realized by $N_i$ different configurations. The probability of a single configuration (represented by a given set of $8$ bits) is given by $p_i=1/N_i$, thus the shape information is given by

\begin{equation}
\label{Shannon-e}
S_i = -K \sum_{l=1}^{N_i}{p_l \ln{p_l}} = -K N_i  {\frac{1}{N_i}} \ln{{\frac{1}{N_i}}} = K \ln{N_i}
\end{equation}

This is same quantity that we have called shape entropy and thus we can interpret the shape entropy as a measure of the information content of a given shape. It is interesting to note that the amount of information differs from class to class. The maximum information is embedded into a random shape, meaning with this a shape whose {\em class} is populated by the whole configuration space: $S_{max} = K \ln{N}$.
In our four particle example, if we assume $K=\log_2e$ and the base $2$ for the $\log$ function, we obtain $S_{max} =7.40$ $bits$. Accordingly, the information in any shape belonging to class $i=2$ and $i=3$ is $S_i=K \log_2(3)=2.29$ $bits$, while the shape information in any shape belonging to class $i=1$ and $i=4$ is just $S_i=K \log_2(1)=0$ $bits$.

Now we proceed by expliciting the connection between a shape change and the minimum energy requirement. In fact, once we have the information associated with each shape, following Landauer\cite{9} and Bennet\cite{10} we deduce that any shape dynamics that involves a change in the information content must play a physical role as well. Specifically, for an isolated system a shape change that implies an increase of shape entropy should come at expenses of a corresponding decrease of the free energy\cite{11}. On the other hand, if we want to perform a shape modification that implies a net shape entropy change of $\Delta S < 0$, this requires a minimum of energy to be dissipated during the transformation equal to $Q = - k_B T \Delta S$. Based on this reasoning we can introduce a special version of the second principle of the thermodynamic in the form: \emph{No process is possible whose sole result is the change of shape of a physical system from a shape of greater shape entropy to a shape of smaller shape entropy}.

We note that the concept of {\em shape entropy} has been already proposed in the literature, although with different definitions in different contexts \cite{2,3,4,5,6}. More relevant to our topics, the role of thermodynamic entropy in stabilizing the shape of Mithocondria has been recently invoked in\cite{7}.  We note that our definition in (\ref{Shape-e}) of shape entropy is in agreement with thermodynamic entropy production along a stochastic trajectory as introduced by Seifert\cite{Seifert} and by Esposito e Van der Broeck\cite{Esposito}. There, the subjected is treated within the framework of the so-called \emph{stochastic thermodynamics} where the entropy variation for a change between two configurations is expressed by\cite{Esposito} $\Delta S=K \ln{(P_m / P_{\bar m})}$.
Where $P_m$ is the probability of the forward trajectory that takes the system from the initial state to the final state and $P_{\bar m}$ is the probability of the reverse
trajectory. In (\ref{Shape-e}) we have:

\begin{equation}
\label{SEsposito-noi}
\Delta S= S_{ij} - S_{rt}  = K  \ln{\frac{N_i}{N_r}} = K \ln{\frac{P_{ij}}{P_{rt}}}
\end{equation}

Where $P_{ij}$ is the forward probability (toward the final state $s_{ij}$) and $P_{rt}$ is clearly the backward probability (toward the initial state $s_{rt}$).

In the remaining part of this letter we discuss the implications of these results in the study of low dissipation physical systems with specific reference to the design of logic switches. These devices have been recently the focus of a great research effort\cite{aa, ab} aimed at decreasing the amount of energy dissipated into heat during a switching event. In fact according to the International Technology Roadmap for Semiconductors we are facing the end of the so-called Moore's law due to the approaching of the limit imposed by the physics of switches: approx $3 k_B T \ln(2)$ of energy per switching event\cite{aa}. Such a limit is due to what is believed to be the minimum amount of energy required to operate a CMOS Field-Effect switch, the building block of present computing devices. In order to overcome such limit the Semiconductor Industry Association of US has launched a search for candidates capable of replacing the CMOS switches. In this perspective nanoscale mechanical switches are currently under investigation with specific attention to their dissipative properties. We propose to address the functioning of a mechanical logic switch by interpreting the switch event as a shape-change process. In fact in any mechanical switch the change from a given logic value (e.g. $0$) to the other (e.g. $1$) comes necessarily with a shape change of some sort and thus we can compute the minimum energy required by such a change based on our previous analysis of the shape entropy change. At difference with the CMOS case, here we are not constrained by specific requirements on geometrical dimensions. If we want to operate the switch in both directions the necessary condition for a zero energy switch is clearly that the the shape entropy change during the switch is zero as well.  Thus the prescription for realizing a zero-power logic switch is to design a switch event between two shapes characterized by the same shape entropy. We point out that the brief analysis presented here is not limited to mechanical switches but it can be easily extended to more general switches where the configuration space instead of being positional is a more abstract space and whose dynamics is acted-on by an equilibrium fluctuation, i.e. connected to a thermal reservoir at fixed temperature $T$. For an extended system, if the configuration space energy landscape is known, then the shape entropy can be computed via the statistical partition function.

\begin{figure}[b]
\includegraphics*[width=9cm]{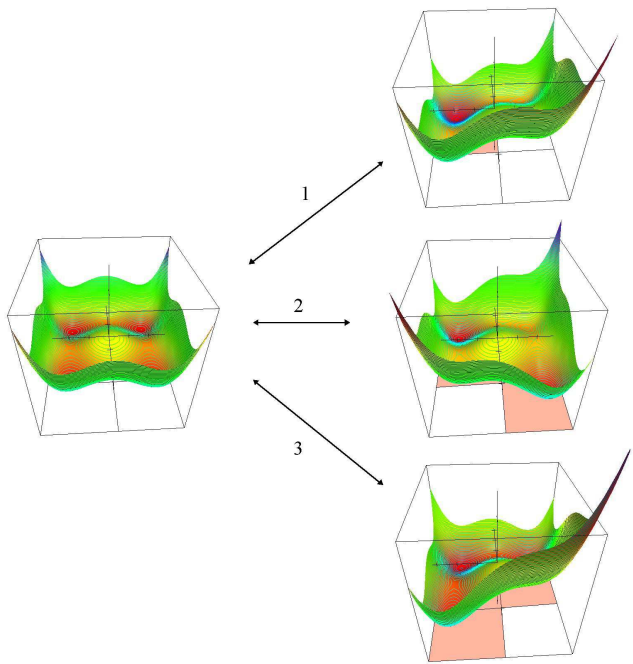}
\caption{Left: unknown status of the potential landscape in the absence of an external force. Right: when a force is applied the potential is changed and the occupation probability is altered. In case 1 the four particles are bound to occupy one single minimum. This is equivalent to resetting the shape of our object to $s_{11}$. Analogous resetting transformation is taking place in 2 (reset to shape $s_{25}$) and in 3 (reset to shape $s_{34}$).}
\label{F2} 
\end{figure}

In order to test our claims on the physical role of shape change, we discuss here some quantitative predictions and propose guidelines for experimental tests in order to verify such predictions. 
Let's consider the shape changes described in Figure 2. In case 1 a static force is applied to the system in order to alter the depth of the potential wells, resulting in a higher probability of occupation for the deeper well. This is equivalent to resetting the shape of our object to $s_{11}1$. In this case the change in entropy can be roughly estimated as follows. Before resetting, the system has an unknown shape. Its shape entropy is thus $S_{max} =  7.40$ $bits$. After the resetting operation the system has shape entropy $S_{1}=0$. Thus the change in shape entropy is $\Delta S =S_{1} -S_{max} = - 7.40$ $bit$ that implies an energy budget of $Q_1 = K_B T \ln(2)\: 7.40$ $J$ to be paid in the process. 
If the experiment is performed at room temperature this amounts to approximately $Q_1 = 2.12\: 10^{-20}$ $J$. Analogously we can compute the energy required for shape change 2 and 3. These result to be $Q_2= Q_3 = 1.47\: \: 10^{-20}$ $J$. In a real experiment the shape change just described can be iterated through a periodic application of the force. If the period of the force $T_p$ is chosen larger than the relaxation time of the associated diffusion process then, due to the entropy change in each cycle, the force will dissipate a power given by $Q/T_p$. The experimental test of our prediction has to be aimed at showing that two different shape changes have different impact on the energy budget of the transformation. In fact, if the shape of an object is just an aesthetic feature without physical content than the energy dissipated during the two transformations will be the same. Conversely the energy will be different. 

What are the best candidates for such a test? In order test our prediction we need to maximize the difference between the dissipated power values corresponding to two different shape changes (e.g. in the example $Q_1/T_p$ and $Q_3/T_p$.). This can be realized by selecting systems that allows for large $N$ differences ($Q$ scales with $\log N$) and small $T_p$. These conditions are best met in nanoscale systems where damping properties play an important role\cite{13} and where, despite recent experimental and theoretical results, the physical mechanism that lies behind energy dissipation in many cases is not completely understood\cite{14,15,16,17}. 

We acknowledge financial support from European Commission (FPVII, G.A. no: 256959, NANOPOWER and no: 270005, ZEROPOWER) and ONRG Grant N 00014-11-1-0695).


\begin{references}

\bibitem[*]{lgemail} e-mail: luca.gammaitoni@pg.infn.it

\bibitem{1} For a reference book see C. W. Gardiner, Handbook of Stochastic Methods, Springer, 2004.

\bibitem{8} C. E. Shannon,  Bell Sys. Tech. Journ., vol. 27, pp. 379-423 and 623-656, 1948.

\bibitem{9} R. Landauer, IBM J. Res. Dev. 5, 183-191 (1961).

\bibitem{10} C. H. Bennett, Int. J. Theor. Phys. 21, 905-940 (1982).

\bibitem{11} S. Toyabe, T. Sagawa, M. Ueda et al., Nature Phys. 6, 12, 988-992 (2010).

\bibitem{2} Louis A. Oddo,  Proc. SPIE 1623, 91 (1992).

\bibitem{3} D. L. Page, A. Koschan, S. Sukumar, B. Abidi, and M. Abidi, Proc. of the Int. Conf. on Image Processing, Vol. I, pp. 229-232, Barcelona, 2003.

\bibitem{4} H.J. Hofmann, J. Sediment. Res. A64, 916Ð920. 1994;  J.P. LeRoux,  Sedimentary Geology  101, 15-20, 1996.

\bibitem{5} F. Wang et al., Proc. ICCV 2003, 0-7695-1950-4/03. 2003.

\bibitem{6} J. Costa and A. O. Hero, in Statistics and analysis of shapes, Eds. H. Krim and T. Yezzi, Birkhauser, pp. 231-252, 2006.

\bibitem{7} M. Ghochani et al., Biophysical Journal V 99 3244Ð3254, 2010.

\bibitem{Seifert} U. Seifert, Phys. Rev. Lett., 95, 040602 (2005).

\bibitem{Esposito} M. Esposito, C.Van den Broeck, Phys. Rev. Lett., 104, 090601 (2010).

\bibitem{aa}	K. Bernstein, R.K. Cavin, W. Porod, A. Seabaugh, J. Welser, Proc. of the IEEE , vol.98, no.12, pp.2169-2184, 2010. 

\bibitem{ab}	 A. M. Ionescu, H. Riel, Nature,  479,   329Ð337, 2011.

\bibitem{13}	See e.g. M. Eichenfield, R. Camacho, J. Chan, K. J. Vahala, and O. Painter, Optomechanical crystals, Nature 459, 550Ð556 (2009).

\bibitem{14}	S. S. Verbridge, J. M. Parpia, R. B. Reichenbach, L. M. Bellan, and H. G. Craighead, Journal of Applied Physics, 99, 124 304 (2006).

\bibitem{15}	A. K. Huettel, G. A. Steele, B. Witkamp, M. Poot, L. P. Kouwenhoven, and H. S. J. van der Zant, Nano Letters, 9, 2547Ð2552 (2009).

\bibitem{16}	A. Eichler, J.Moser, J. Chaste, M. Zdrojek, I. Wilson-Rae and A. Bachtold,  Nature Nanotech. 6, 339Ð342 (2011).

\bibitem{17}	Q. P. Unterreithmeier, T. Faust, and J. P. Kotthaus,  Phys. Rev. Lett. 105, 027205 (2010).

\end{references}
\end{document}